\documentclass[iop]{emulateapj}
\usepackage{amssymb,amsmath}
\usepackage[colorlinks=true,linkcolor=blue,anchorcolor=blue,citecolor=blue,urlcolor=blue]{hyperref}
\usepackage{ulem}
\usepackage{color}
\usepackage{natbib}
\usepackage[hang,flushmargin]{footmisc} 
\slugcomment{}

\shorttitle{The Environment of Massive Quiescent Compact Galaxies at $0.1<z<0.4$ in the COSMOS Field}
\shortauthors{Damjanov et al.}

\begin{document}

\title{The Environment of Massive Quiescent Compact Galaxies at $0.1<z<0.4$ in the COSMOS Field}

\author{Ivana Damjanov\altaffilmark{1}, H. Jabran Zahid\altaffilmark{2}, Margaret J. Geller\altaffilmark{2}, Ho Seong Hwang\altaffilmark{3}}
\altaffiltext{1}{Harvard-Smithsonian Center for Astrophysics, 60 Garden Street, Cambridge, MA 02138;  \href{mailto:idamjanov@cfa.harvard.edu}{idamjanov@cfa.harvard.edu}}
\altaffiltext{2}{Smithsonian Astrophysical Observatory, 60 Garden St., Cambridge, MA 02138}
\altaffiltext{3}{School of Physics, Korea Institute for Advanced Study, 85 Hoegiro, Dongdaemun-gu, Seoul 130-722, Republic of Korea}

\begin{abstract}

We use Hectospec mounted on the 6.5-meter MMT to carry out a redshift survey of red ($r-i>0.2$, $g-r>0.8$, $r<21.3$) galaxies in the COSMOS field to measure the environments of massive compact quiescent galaxies at intermediate redshift. The $>90\%$ complete  magnitude limited survey includes redshifts  for 1766 red galaxies with $r < 20.8$ covering the central square degree of the field; 65\% of the redshifts in this sample are new. We select a complete magnitude limited quiescent sample based on the rest-frame $UVJ$ colors. When the density distribution is sampled on a scale of 2~Mpc massive compact galaxies inhabit systematically denser regions than the parent quiescent galaxy population. Non-compact quiescent galaxies with the same stellar masses as their compact counterparts populate a similar distribution of environments. Thus the massive nature of quiescent compacts accounts for the environment dependence and appears fundamental to their history.

\end{abstract}


\section{Introduction}

Environment plays a significant role in galaxy mass assembly. Internal galaxy properties, luminosity and morphology, are correlated with the local galaxy density at $z\sim0$ \citep[e.g.,][]{Davis1976, Park1994, Park2007}. This correlation extends to $z\sim1$ and to the relation between stellar mass and environment \citep[e.g.,][and references therein]{Darvish2015}. 

Massive compact quiescent galaxies are extreme probes for models of massive galaxy formation and evolution. These systems were first discovered at high redshift ($z>1$) where massive quiescent galaxies are on average several times smaller (i.e., have higher surface stellar mass density) than typical passive systems of similar stellar mass at $z\sim0$ \citep[e.g.,][]{Daddi2005, Longhetti2007,Trujillo2007, Toft2007, Zirm2007, Cimatti2008, vanDokkum2008, Buitrago2008, Damjanov2009, Damjanov2011, McLure2012, vandeSande2013, Belli2014}. Recently discovered ultra-diffuse galaxies found in the Coma cluster \citep{vanDokkum2015, Koda2015} are at the other extreme of the surface density distribution. These objects are passively evolving, Milky Way sized galaxies with extremely low surface brightness ($25-28$~mag~arcsec$^{-2}$ in $R-$band). The exploration of the links between both the extraordinarily dense and the unusually diffuse galaxies and their environments should provide further clues to the driving mechanisms of galaxy assembly. Using recent cosmological simulations, \citet{Stringer2015} traced the location of massive compact systems evolving into galaxies similar to NGC 1277, the first known local relic of a massive compact galaxy formed at $z\gtrsim2$ \citep{VandenBosch2012, Trujillo2014}. The results suggest that a large fraction of  massive compact systems are substructures of more massive groups and clusters of galaxies at $0<z<2$; i.e, massive compacts preferentially occupy denser regions. These and other results \citep[e.g.,][]{Shankar2014} suggest that, indeed, environment may have an important role in the formation, preservation and structural evolution of massive compact systems in the intermediate redshift regime.  

If  size growth is efficient, massive compact galaxies constitute a significant fraction of the quiescent population only at high redshift  \citep[e.g.,][]{Huertas-Company2015, Wellons2015a, Wellons2015b} where dense spectroscopic surveys, essential for characterization of the galaxy density field \citep[e.g.,][]{Geller2015}, are not feasible. The recent identification of a substantial number of these systems at intermediate redshift \citep[e.g.,][]{Damjanov2015} makes the study of their environments over the last $4-5$~Gyr possible.

Estimates of the number density evolution of compact massive galaxies  range from a dramatic change in the number density between $z\sim2$ and the local universe \citep[e.g.,][]{Trujillo2009, Taylor2010, vanderWel2014}  to a very mild evolution \citep[e.g.,][]{Poggianti2013a, Carollo2013, Damjanov2014, Damjanov2015, Saulder2015, Tortora2015}. \citet{Damjanov2015} use a spectroscopic sample of intermediate-redshift ($0.2<z<0.8$) quiescent galaxies drawn from the COSMOS survey to show that the number density of massive compact galaxies is consistent with no evolution in this redshift range. The abundance of intermediate-redshift compacts is similar to their number density at $z>1$. 

The abundance of massive compact galaxies at intermediate redshifts provides an opportunity to study the evolution of these systems in much more detail than is currently possible at higher redshift. Large and dense spectroscopic surveys can provide  substantial samples of compact systems along with central velocity dispersion measurements for individual objects \citep[e.g.,][Zahid et al. 2015 in prep.]{Fabricant2013, Monna2014, Zahid2015}. Large redshift surveys covering the range $0.1 <z< 0.8$ can probe volumes encompassing environments ranging from galaxy groups to the supercluster - void network. A dense redshift survey (i.e., a  $>90\%$ complete spectroscopic survey given the selection) enables evaluation of the quiescent galaxy density field around compact quiescent systems on a variety of scales.

Several studies have characterized the evolution of structure in the COSMOS field \citep[e.g,][]{Scoville2007, Scoville2013, Kovac2010} and investigated the links between internal properties of COSMOS galaxies and their environments \citep[e.g.,][]{Guzzo2007, Cassata2007, Capak2007, Ideue2009, Meneux2009, Bolzonella2010, George2011, Kovac2013, Darvish2015}. Here we use a new, dense and complete, redshift survey to explore  the relation between intrinsic and environmental properties of massive compact galaxies in COSMOS at $0.1<z<0.4$.

We combine a spectroscopic survey carried out with Hectospec on the 6.5-m MMT with publicly available data to construct a complete magnitude limited sample of $880$ intermediate redshift quiescent galaxies in the COSMOS field. The survey densely samples the distribution of massive quiescent galaxies at intermediate redshift. 

We compare the galaxy stellar mass density distribution around massive compact galaxies with  the environments of non-compact massive quiescent galaxies. Section~\ref{data} describes the spectroscopic survey and the selection. We describe the technique we use to evaluate the  galaxy stellar mass density field in Section~\ref{method} and our results are in Section~\ref{densityfield}. We explore the relationship between galaxy stellar mass and environment in Section~\ref{mass}. We discuss the results in Section~\ref{comparison} and conclude in Section~\ref{sum}. We adopt $\Omega_{\Lambda}=0.7$, $\Omega_{M} = 0.3$, and $H_0 = 70$~km~s$^{-1}$~Mpc$^{-1}$ cosmology.       

\section{The Data}\label{data}

\begin{figure}
\begin{centering}
\hspace*{-0.3in}
\includegraphics[scale=0.27]{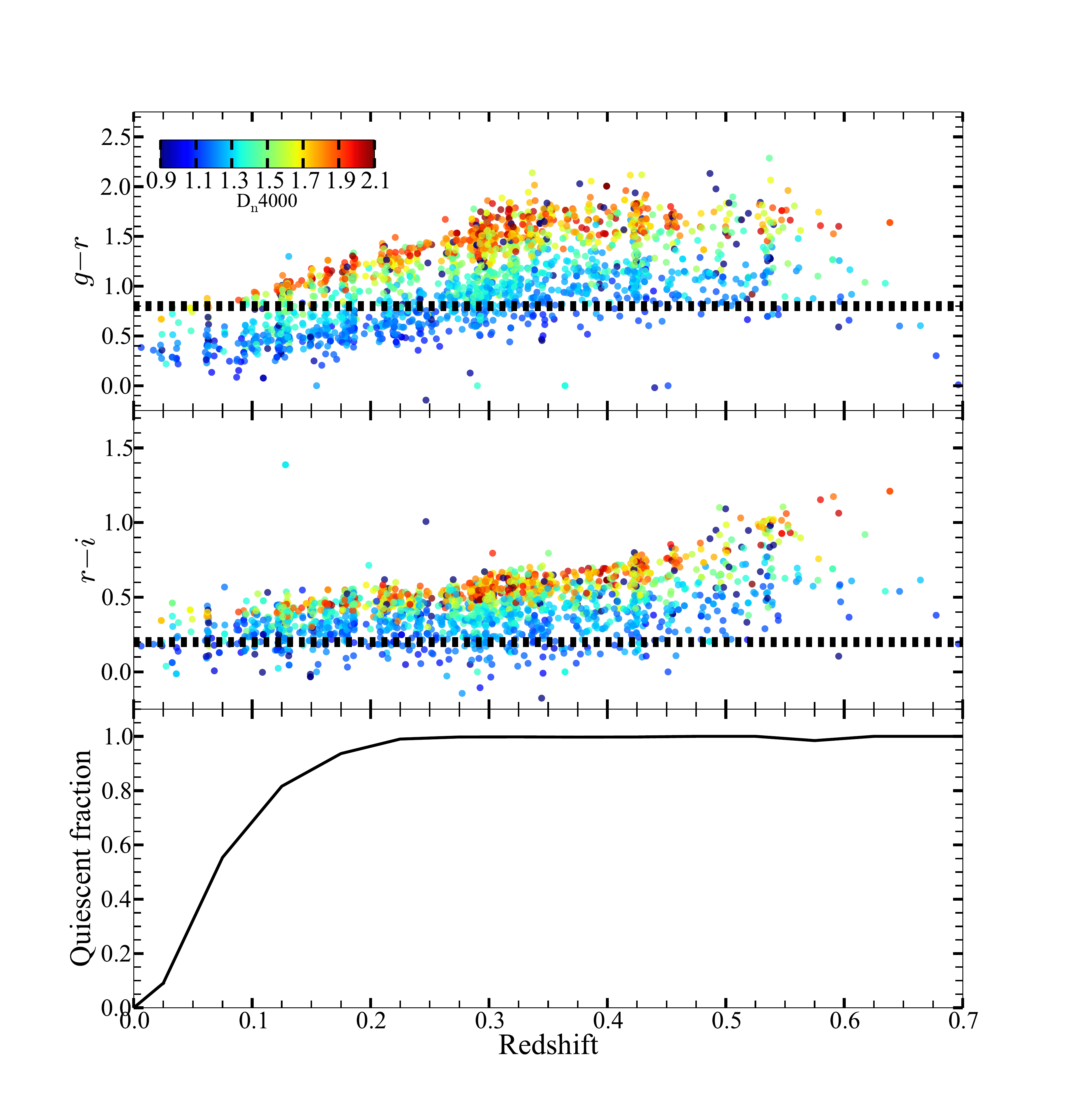}
\caption{Color selection of the hCOSMOS+zCOSMOS survey: two upper panels show the observed $g-r$  and $r-i$ colors of SHELS galaxies \citep{Geller2014} as functions of redshift. Dashed lines label our color selection: $g-r>0.8$ and $r-i>0.2$. Individual points are color-coded by the indicator of galaxy quiescence, D$_\mathrm{n}4000$. The third panel shows the fraction of quiescent  (D$_\mathrm{n}4000>1.44$) SHELS systems with $gri$ colors of  hCOSMOS+zCOSMOS galaxies.\label{f1}}
\end{centering}
\end{figure}

We use Hectospec \citep{Fabricant1998, Fabricant2005} mounted on the 6.5-meter MMT to measure redshifts for galaxies drawn from the UltraVISTA catalog\footnote{\url{http://www.strw.leidenuniv.nl/galaxyevolution/ULTRAVISTA/Ultravista/K-selected.html}}\citep{Muzzin2013a}. We apply a color selection ($r-i>0.2$, $g-r>0.8$) and a magnitude limit of $r<21.3$. We refer to the Hectospec survey of the COSMOS field as hCOSMOS hereafter. 

We measure redshifts for galaxies without spectroscopy in the existing public databases (zCOSMOS redshift survey, \citealt{Lilly2007, Knobel2012}; SDSS DR12, \citealt{Alam2015}). Our goal is to increase the redshift survey density in the redshift range $0.1<z<0.4$ as a basis for investigating the environments of massive quiescent compact galaxies.

With the inclusion of Hectospec spectroscopy, we construct a $90\%$ complete sample of 1766 galaxies with ($r-i>0.2$, $g-r>0.8$) and $r<20.8$. This sample is larger and more uniform than samples that have been considered previously at $0.1<z<0.4$ \citep[e.g.,][]{Maltby2010, Huertas-Company2013a, Lani2013, Kelkar2015}.

To define observed color cuts which provide a complete magnitude limited sample of intermediate-redshift quiescent galaxies we examine the Smithsonian Hectospec Lensing Survey \citep[SHELS;][]{Geller2014}. SHELS is complete to $R=20.6$ with no color selection. 

Figure~\ref{f1} shows the effect that our color selection has on the SHELS data for the limiting magnitude $R=20.6$. The first two panels show $g-r$ and $r-i$ colors of SHELS $R<20.6$ galaxies in the $0<z<0.7$ redshift range. Symbols are color-coded by D$_\mathrm{n}4000$ values measured from the SHELS Hectospec spectra. The $g-r>0.8$ cut effectively provides a sample of quiescent system at $z\gtrsim0.1$; $r-i>0.2$ color selection minimizes the contamination by foreground objects (at $z<0.1$). The bottom panel of Figure~\ref{f1} demonstrates that the fraction of quiescent galaxies (with D$_\mathrm{n}4000>1.44$, see Section~\ref{qsamp}) in the galaxy sample with red $gir$ colors increases steeply between $z=0$ and $z=0.1$ and exceeds 90\% at $z\sim0.15$.

The magnitude and color cuts we apply efficiently select a complete magnitude limited sample of quiescent galaxies in the redshift range $0.1\lesssim z \lesssim 0.4$ (in a small fraction of the volume completeness is in the 80-90\% range). Because of the broad color selection our sample also contains some star-forming galaxies. Thus we apply additional selection criteria to construct a quiescent subsample (Section~\ref{qsamp}).  We base our analysis on the quiescent subset of 880 galaxies.

\subsection {Spectroscopy}

The 1 square degree field of view of Hectospec is well matched to the size of the COSMOS field. For galaxies with $r<21.3$, a typical integration time of 1~h yields a redshift \citet{Fabricant2008}. Because our intermediate-redshift sample is predominantly quiescent, redshift determination is based on the prominent absorption features (e.g, Ca H+K, Balmer series, G-band) and on the position of 4000~\AA\ break. Strong emission lines (e.g, [OII] and/or H$\alpha$) determine redshifts of emission line galaxies. We prioritize field positions around the center ($\alpha_{2000}=\, $10:00:28.6, $\delta_{2000}=\, $+02:12:21.00) of the COSMOS field. We also prioritize galaxies according to their apparent $r-$band magnitudes and use the \citet{Roll1998} software to optimize the fiber positions. We obtained 10 fields in varying conditions in February and April 2015. Damjanov et al. (2015, in prep) will include a detailed description of this dataset.

We reduce the Hectospec data with the \citet{Mink2007} Hectospec pipeline and derive redshifts with HSRED v2.0, developed by the Telescope Data Center and based on the pipeline originally developed by Richard Cool\footnote{\url{http://www.mmto.org/book/export/html/55}}. Repeat observations show that the typical error in the Hectospec redshift is 48 km  s$^{-1}$ \citep[normalized by $(1 + z)$;][]{Geller2014}.

The pipeline returns the  $r$-value \citep{Tonry1979} as a measure of the quality of the redshift. We also visually inspect all of the spectra; we use only those spectra which yield high-quality redshifts (i.e., redshifts based on several reliable spectral features and with $r$-values $\gtrsim3$; see \citealt{Fabricant2005}).

We acquired a total of 2096 new redshifts. Here we use 1160 of these spectra for  galaxies with $r<20.8$ (see Section~\ref{speccomp}) and combine them with other publicly available spectroscopic data in the COSMOS field. Table~\ref{tab1} lists the source and the numbers of redshifts we use. The entire compilation contains $10,314$~galaxies in the redshift range $0\leqslant z<1.35$ with secure spectroscopic redshifts, reliable photometry and GIM2D-based galaxy size measurements from \citet{Sargent2007}\footnote{\url{http://irsa.ipac.caltech.edu/data/COSMOS/tables/morphology/cosmos_morph_zurich_1.0.tbl}}.We use a subsample with high spectroscopic completeness (Section~\ref{speccomp}) in the central region of the COSMOS field to select quiescent galaxies based on their rest-frame colors (Section~\ref{qsamp}) and to probe the environments of quiescent systems at intermediate redshift.

\begin{deluxetable*}{ccccc}
\tabletypesize{\small}
\tablecaption{Properties of the hCOSMOS+zCOSMOS redshift survey\label{tab1}}
\tablewidth{7in}
\tablehead{ \colhead{Sample\tablenotemark{a}} & \multicolumn{4}{c}{Number of spectroscopic sources} \\
\colhead{} & \colhead{Total\tablenotemark{b}} & \colhead{$r<20.8$\tablenotemark{c}} & \colhead{Quiescent\tablenotemark{d}} & \colhead{Compact\tablenotemark{e}} \\
\colhead{} & \colhead{} & \colhead{} & \colhead{($r<20.8$)} & \colhead{($r<20.8$)} \\  
\colhead{(1)} & \colhead{(2)} & \colhead{(3)} & \colhead{(4)} & \colhead{(5)} \\ 
}
\startdata
hCOSMOS & 2096 & 1160 & 565 & 167\\  
zCOSMOS & 16529 & 529 & 254 & 81\\
SDSS & 960 & 77 & 61 & 23\\
\hline
&&&&\\
Total & 19585 & 1766 & 880 & 271\\
\enddata
\tablenotetext{a}{References: Damjanov et al. 2015 (in prep, hCOSMOS), \citet[zCOSMOS]{Lilly2007}, \citet[zCOSMOS]{Knobel2012}, \citet[SDSS DR12]{Alam2015}}
\tablenotetext{b}{Number of unique high-quality spectroscopic redshifts in the sample}
\tablenotetext{c}{Number of ($r-i>0.2$, $g-r>0.8$) spectroscopic targets with: 1) apparent magnitude $r<20.8$,  2) reliable redshift, 3) available stellar masses, and 4) measured structural properties in the central 1 sq. degree of the COSMOS field (where spectroscopic completeness is $>90\%$ to the magnitude limit, see Section ~\ref{speccomp}.)}
\tablenotetext{d}{Number of ($r-i>0.2$, $g-r>0.8$, $r<20.8$) spectroscopic targets with:  1) reliable redshift, 2) available stellar masses, 3) measured structural properties in the central 1 sq. degree of the COSMOS field, and 4) rest-frame colors of quiescent galaxy population (see Section ~\ref{qsamp}.)}
\tablenotetext{e}{Number of ($r-i>0.2$, $g-r>0.8$, $r<20.8$) spectroscopic targets with:  1) reliable redshift, 2) available stellar masses, 3) measured structural properties in the central 1 sq. degree of the COSMOS field, 4) rest-frame colors of quiescent galaxy population, and 5) compact structure defined by Equation~\ref{eq:def}}
\end{deluxetable*}

\subsection{Completeness}\label{speccomp}

\begin{figure}
\begin{centering}
%\hspace*{-0.15in}
\includegraphics[scale=0.25]{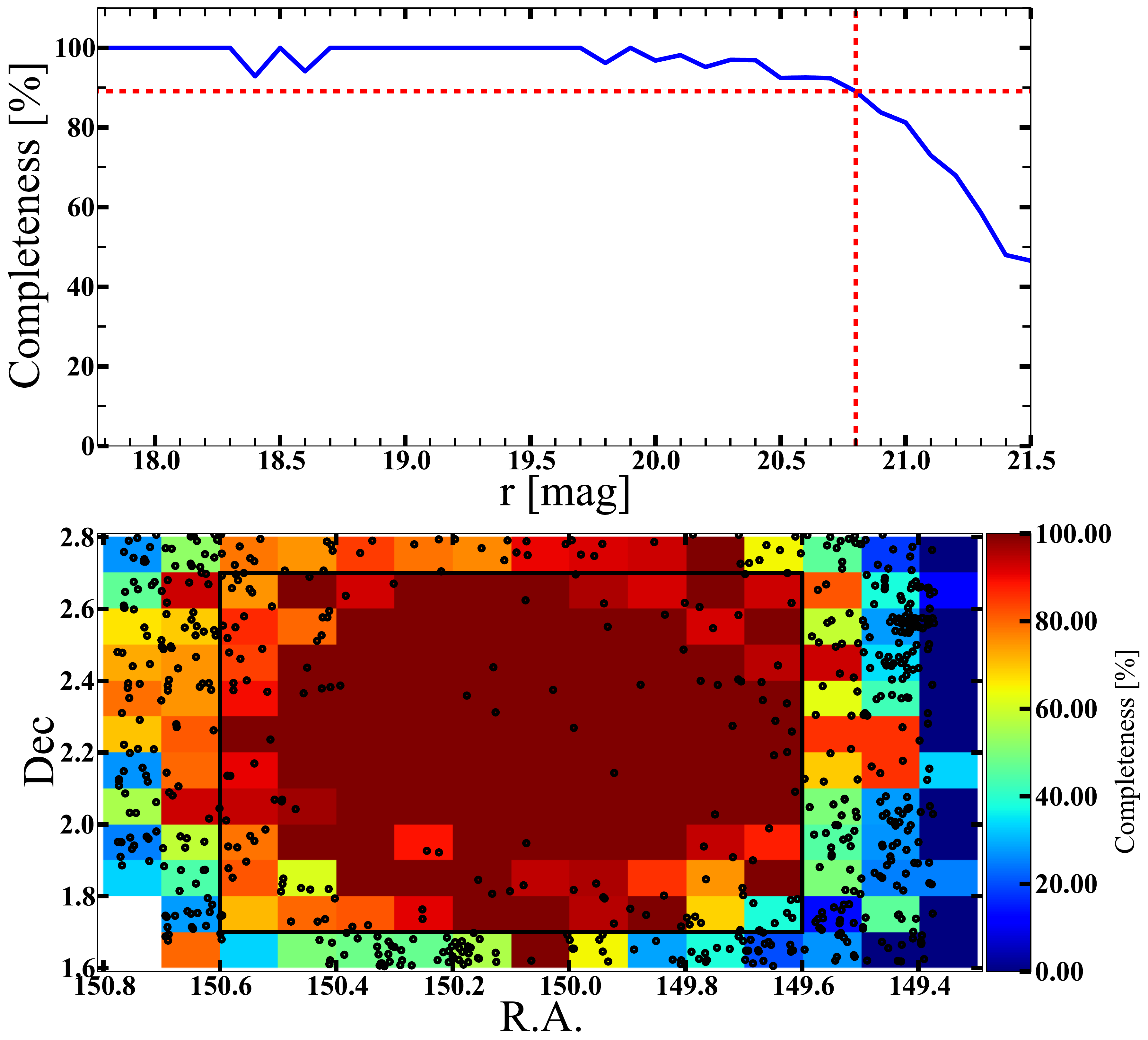}
\caption{Completeness of the hCOSMOS+zCOSMOS redshift survey. The upper panel shows the differential completeness as a function of $r-$band magnitude (blue). The vertical red line at $r = 20.8$ marks the 90\% differential completeness limit.  The lower panel shows the spectroscopic completeness in $6\arcmin\times6\arcmin$ bins for galaxies with $r<20.8$. The black rectangle outlines the central 1 square degree of the field where the differential spectroscopic completeness is $>90\%$ for quiescent galaxies with $r<20.8$. Black points indicate targets in the photometric ULTRAVISTA sample \citep{Muzzin2013a} without a measured redshift.\label{f2}}
\end{centering}
\end{figure}

We examine the completeness of the redshift survey as a function of position on the sky and as a function of limiting $r$ magnitude. We focus on the central square degree where the hCOSMOS+zCOSMOS survey is the most dense (black rectangle in the lower panel of Figure~\ref{f2}).

We use the photometric UltraVISTA catalog \citep{Muzzin2013a} to define the spectroscopic completeness. The upper panel of Figure~\ref{f2} shows the differential ratio of spectroscopic to photometric targets with $r-i>0.2$ and $g-r>0.8$ as a function of $r-$band magnitude.  The fraction of galaxies with a reliable spectroscopic redshift exceeds 90\% for $r < 20.8$. To this limit, the sample includes 1766 red galaxies with

\begin{itemize}   
\item a spectroscopic redshift $0.1<z\leqslant0.8$;
\item stellar mass $10^{9}\, M_\sun<M_\star\leqslant6\times10^{11}\, M_\sun$ \citep{Muzzin2013a};
\item circularized effective (or half-light) radius ${0.2\, \mathrm{kpc}<R_{e,c}<33\, \mathrm{kpc}}$ \citep{Sargent2007}.
\end{itemize}

\noindent Redshifts for a majority of these objects (65\%, see Table~\ref{tab1}) are from the hCOSMOS survey. Measured radii for all objects exceed the average width of the Point Spread Function (PSF) reported for the HST ACS mosaic of the COSMOS field \citep[$0\farcs095$,][] {Koekemoer2007}. \citet{Muzzin2013a} determine stellar population parameters by fitting galaxy spectral energy distributions (SEDs) using the FAST code \citep{Kriek2009}. We  use stellar masses based on the best-fit \citet{Bruzual2003} models with solar metallicity, a \citet{Chabrier2003} Initial Mass Function, \citet{Calzetti2000} dust extinction law, and exponentially declining star formation histories.

The majority of ($r-i>0.2$, $g-r>0.8$) color-selected galaxies with $r<20.8$ and without spectroscopic (hCOSMOS, zCOSMOS, or SDSS) redshifts are distributed towards the edges of the COSMOS field (lower panel of Figure~\ref{f2}). In the central 1 sq. degree of  the field (black rectangle) there are 135 photometric targets without a spectroscopic redshift ($\lesssim0.8\%$ of the spectroscopic sample). A majority of the unobserved photometric targets ($\sim60\%$) have $r>20.4$, i.e., their magnitude distribution is skewed toward the limiting magnitude. The high completeness results from the hCOSMOS survey design. 

\subsection{The Quiescent Galaxy Sample}\label{qsamp}

\begin{figure*}
\begin{centering}
\includegraphics[scale=0.45]{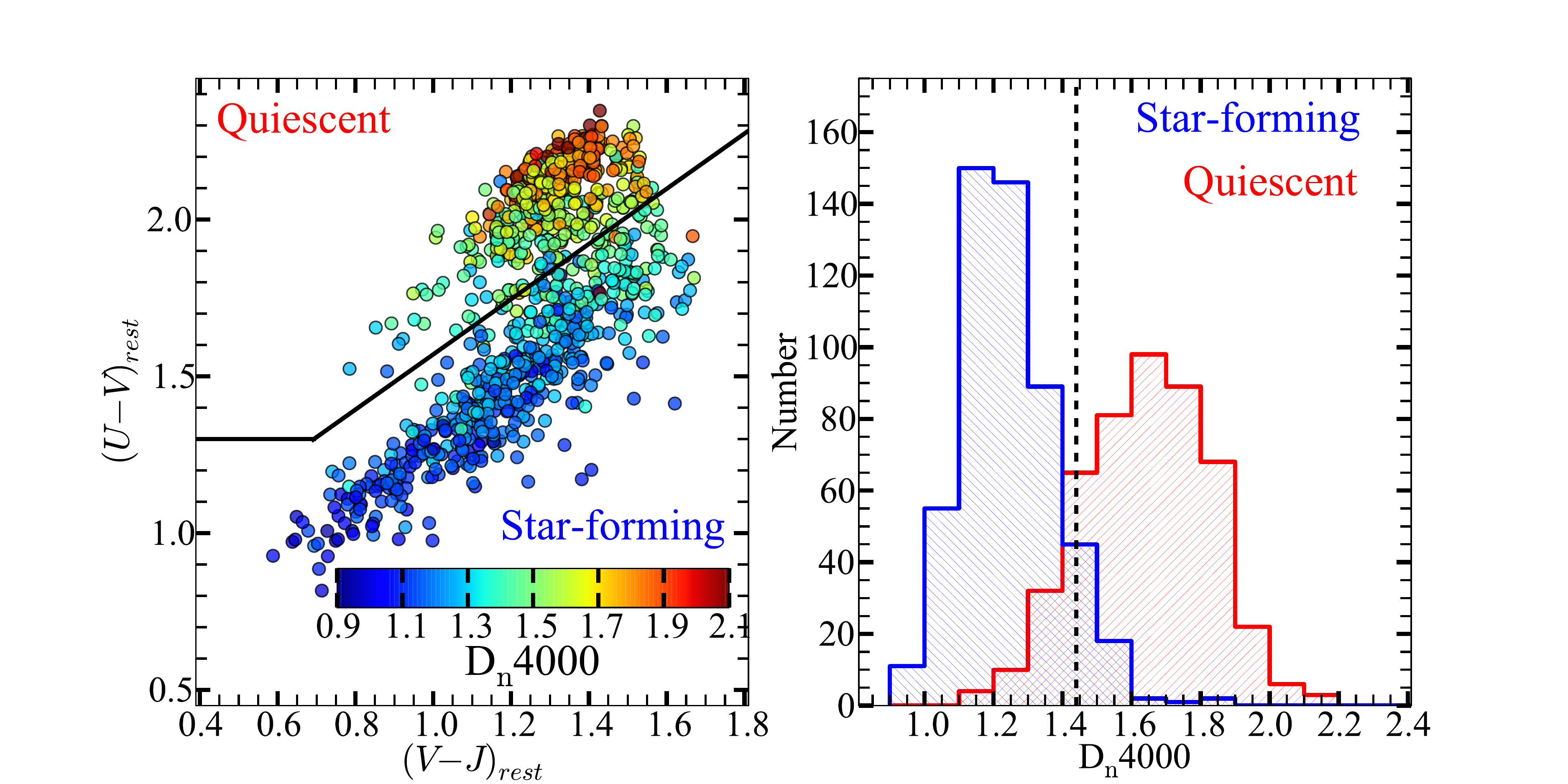}
\caption{Left panel: Rest-frame $UVJ$ color-color diagram for hCOSMOS and SDSS galaxies with measured D$_{\mathrm{n}}4000$. Color-coding indicates D$_{\mathrm{n}}4000$. The separation between quiescent and star-forming galaxies (solid black line) is from \citet[equations~\ref{eq:qs1}~and\ref{eq:qs2}]{Williams2009}. Right panel: The distribution of D$_{\mathrm{n}}4000$ for quiescent and star-forming galaxies selected based on their rest-frame $UVJ$ colors. \label{f3}}  
\end{centering}
\end{figure*}

The broad color selection of the hCOSMOS survey provides a complete sample of quiescent galaxies in the redshift range $0.1<z<0.4$. However, at higher redshifts probed by the survey the observed ($r-i>0.2$, $g-r>0.8$) color limits select a broader range of rest-frame colors that extends to the blue. Thus the fraction of star-forming galaxies increases with redshift. We examine the galaxy stellar mass density field around quiescent galaxies using these quiescent systems as tracers of the total stellar mass density in quiescent objects. The first step in our analysis is the extraction of a complete sample of quiescent galaxies.

Combinations of rest-frame colors provide diagnostics for galaxy star-formation activity \citep[e.g.,][]{Williams2009, Ilbert2010}. We use rest-frame $U-V$ and $V-J$ colors from \citet{Muzzin2013a}\footnote{\url{http://www.strw.leidenuniv.nl/galaxyevolution/ULTRAVISTA/Ultravista/K-selected.html}} to separate star-forming and passively evolving galaxies based on the selection criteria for $0<z<0.5$ quiescent galaxies from \citet{Williams2009}:

\begin{subequations}
\begin{alignat}{1}
 U-V&>1.3  \label{eq:qs1}\\
 U-V&>(V-J)\times0.88+0.69 \label{eq:qs2}. 
 \end{alignat}
 \end{subequations} 
 
\noindent This color selected quiescent sample includes 880 intermediate-redshift COSMOS galaxies.

All quiescent galaxy selection criteria produce samples with some contamination from star-forming outliers \citep{Moresco2013}. The hCOSMOS and SDSS spectra span the range $\sim3700-9000$~\AA . Thus we can use D$_\mathrm{n}4000$ \citep[as defined in][]{Balogh1999} to test the contamination of the rest-frame color selected quiescent sample.     

The D$_\mathrm{n}4000$ spectral index is an indicator of galaxy quiescence: D$_\mathrm{n}4000=1.44$ effectively separates absorption-line from the emission-line systems \citep[][]{Woods2010}. We follow \citet{Fabricant2008} procedure to measure D$_\mathrm{n}4000$ for the hCOSMOS galaxy sample and combine it with measurements available for the SDSS/COSMOS spectroscopic targets \citep{Kauffmann2003}. \citet{Fabricant2008} demonstrate that D$_\mathrm{n}4000$ measured from the Hectospec galaxy spectra agree remarkably well with the values based on SDSS spectra. We compile D$_\mathrm{n}4000$ measurements with small errors ($<0.09$ of the index value, or $<2\times$ typical error; \citealt{Fabricant2008}) for $57\%$ of the COSMOS dataset. We use this subsample to examine the contamination of the galaxy sample drawn from their rest-frame colors. 

Figure~\ref{f3} shows the position of hCOSMOS and SDSS/COSMOS galaxies as a function of their rest-frame ($U-V$, $V-J$) colors (left panel) and the distribution of D$_\mathrm{n}4000$ for the star-forming and quiescent subsamples defined by the rest-frame $UVJ$ selection (\citealt{Williams2009}; Equations~\ref{eq:qs1}~and~\ref{eq:qs2}; right panel).  In the passive color-selected subsample $\sim15\%$ of galaxies have D$_\mathrm{n}4000<1.44$; $\sim10\%$ of galaxies with rest-frame colors indicating star formation have D$_\mathrm{n}4000>1.44$. Of course some galaxies with large D$_\mathrm{n}4000$ are star-forming and others with D$_\mathrm{n}4000<1.44$ are quiescent \citep{Woods2010}. Thus this comparison, like all other classifiers, is only indicative of the possible contamination.

\citet{Moresco2013} select  a set of quiescent galaxy samples at  $z<0.5$ based on a variety of selection criteria ranging from early-type morphologies to a combination of photometric and spectroscopic properties. The authors demonstrate that in all quiescent samples $21-55\%$  of galaxies have prominent emission lines (the only spectroscopic indicator of star formation used in their study). The contamination level of our rest-frame $UVJ$ color selected sample of $15\%$ (estimated from the distribution of D$_\mathrm{n}4000$ values) is lower than in any of the \citet{Moresco2013} samples tested with spectroscopic indicators.  

 \subsection {The Compact Quiescent Galaxy Sample}\label{comp}
 \begin{figure}
\begin{centering}
\hspace*{-0.15in}
\includegraphics[scale=0.25]{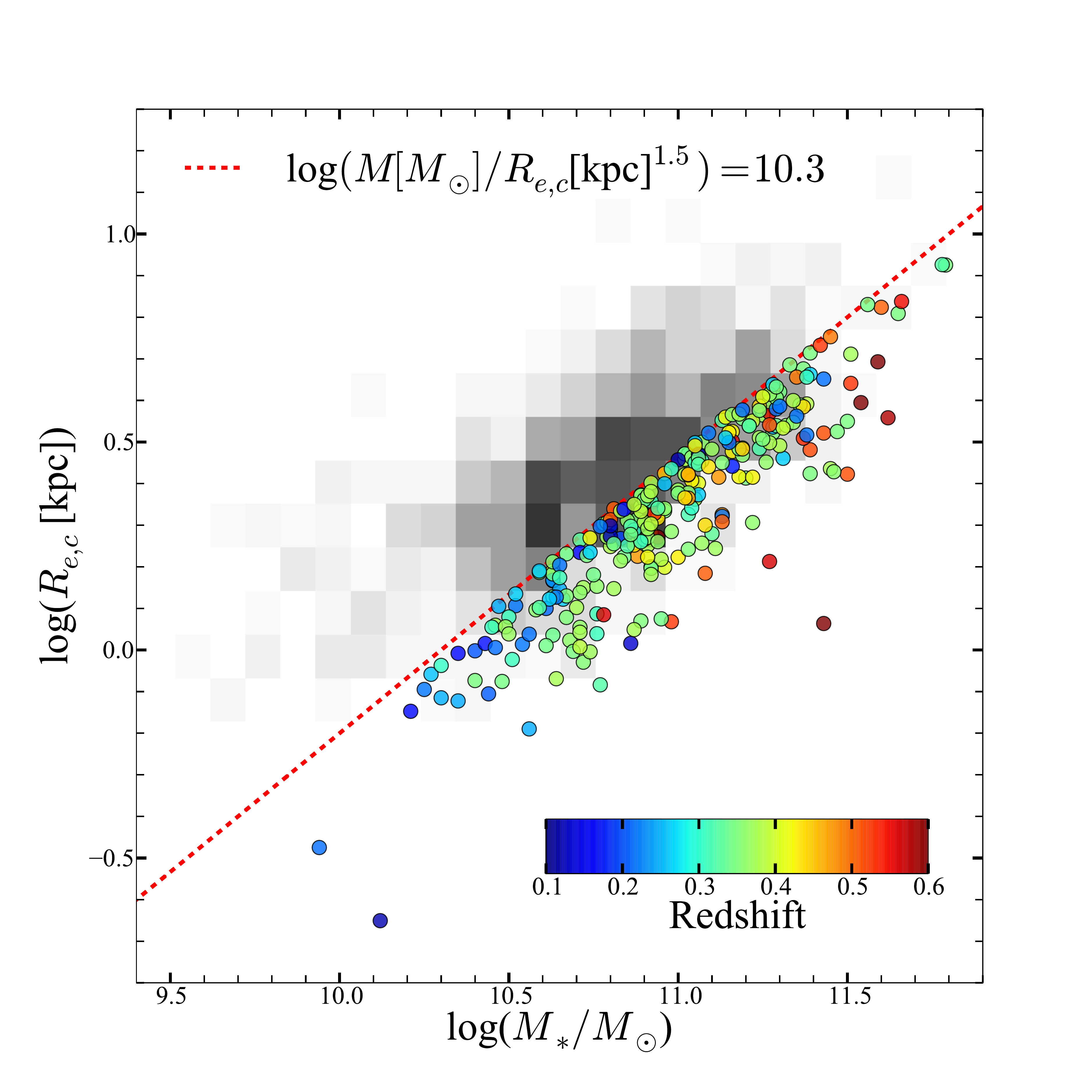}
\caption{Circularized effective radius as a function of stellar mass for the COSMOS compact galaxies (circles color-coded by redshift). The gray two-dimensional histogram represents the distribution of the parent quiescent intermediate-redshift sample. The red dashed line shows the compactness cutoff. \label{f4}}  
\end{centering}
\end{figure}

\begin{figure*}
\begin{centering}
\includegraphics[scale=0.45]{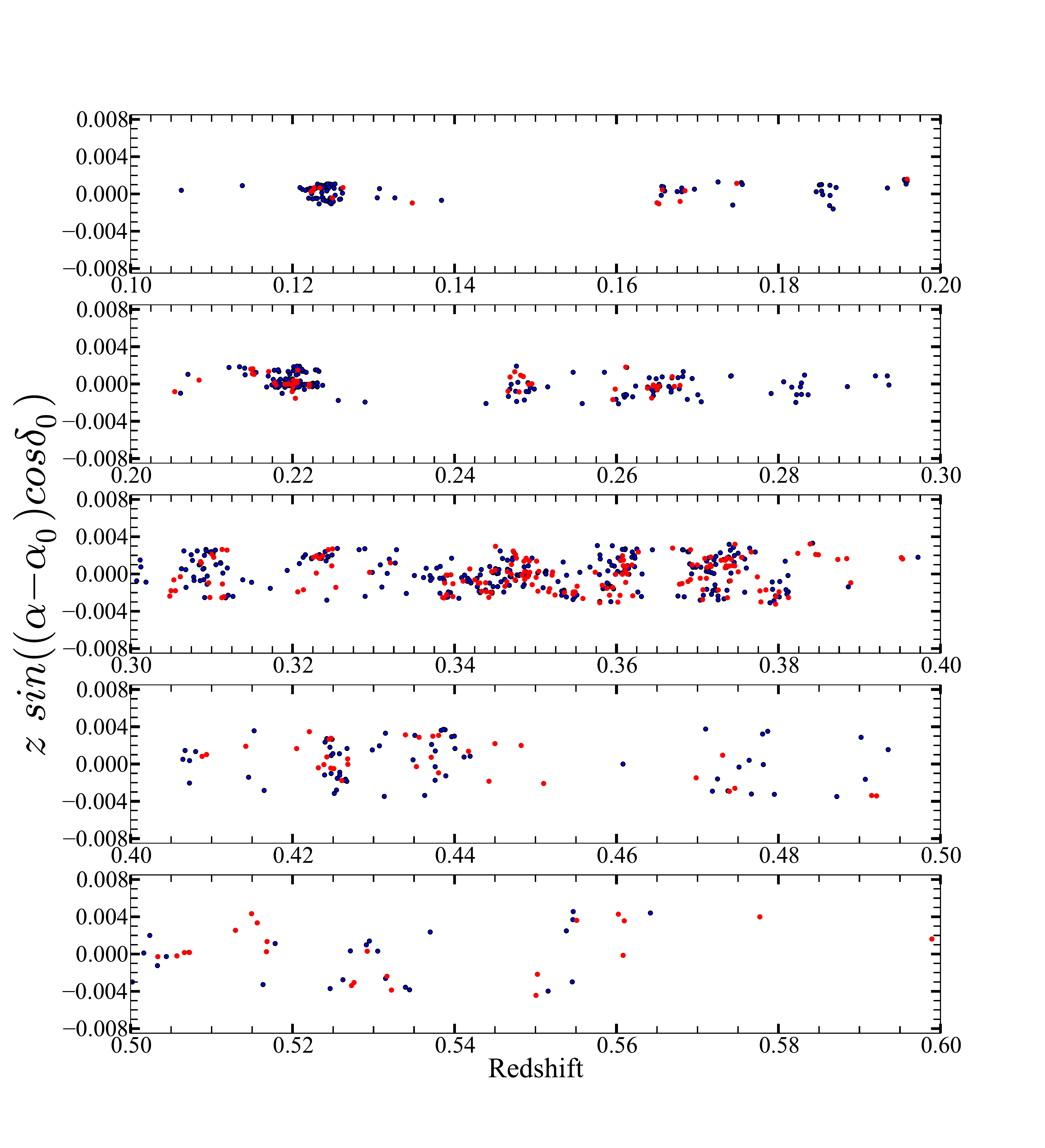}
\caption{Cone diagram for quiescent galaxies in the $r<20.8$ hCOSMOS+zCOSMOS survey projected in R.A.$_\mathrm{2000}$. Red dots indicate massive compact quiescent galaxies \citep{Barro2013}; blue dots indicate other quiescent objects.
\label{f5}}  
\end{centering}
\end{figure*}

There are a variety of definitions for compact galaxies \citep[e.g.,][]{Trujillo2009, Cassata2013, vanderWel2014, Barro2013}. In \citet{Damjanov2015} we show that global physical parameters like the dependence of the space density on redshift are insensitive to the compact galaxy definition. Thus we use a definition that strikes a balance between the selection of the densest massive  systems and the necessity of having a large enough sample for a robust analysis of the environments. We use the pseudo-stellar mass surface density $\Sigma_{1.5}$:

\begin{equation}
\Sigma_{1.5} \equiv log\left(\frac{M_\star[M_\sun]}{\left(R_{e,c}[\mathrm{kpc}]\right)^{1.5}}\right) \label{eq:def},
\end{equation}

\noindent and  a compact galaxy threshold $\Sigma_{1.5}\geqslant10.3$ \citep[e.g.,][]{Barro2013, Poggianti2013a, Damjanov2015}. Figure~\ref{f4} illustrates the position of the selected compact systems (circles colored by redshift) in the size-stellar mass parameter space compared with the distribution of the parent quiescent COSMOS sample (gray two-dimensional histogram). The threshold based on pseudo-stellar mass surface density (Eq.~\ref{eq:def}, red dashed line in Figure~\ref{f4}) produces a sample of 271 massive ($M_\star\gtrsim10^{10}\, M_\sun$) compact red galaxies at $0.1<z\lesssim0.6$. 

Figure~\ref{f5} shows a cone diagram projected along the R.A.$_{\mathrm{2000}}$ direction. Galaxies are color-coded based on their pseudo-stellar mass surface density $\Sigma_{1.5}$ (Eq.~\ref{eq:def}). Red circles represent massive compact galaxies and blue circles show other quiescent galaxies. One clear example of massive compact quiescent systems in a dense region is the set of nine compact systems in the core of a galaxy system (note the extended finger in redshift space) at $z=0.22$. Furthermore, the large number of compact systems at $z\sim0.35$ coincides with a known significant galaxy overdensity in COSMOS \citep[e.g, ][]{Masters2011}.

The bottom panels of Figure~\ref{f5} cover the $0.4\leqslant z<0.6$ redshift interval and thus the largest fraction ($\sim$70\%) of the volume. However, this redshift range contains a small fraction of the quiescent galaxy sample (146 objects or $\sim$17\%). Comparison with the number density of massive quiescent COSMOS galaxy sample from Damjanov et al. 2015 confirms that the hCOSMOS+zCSOMOS galaxy sample  misses $\gtrsim50\%$ of massive ($M_\star>10^{10}\, M_\sun$) quiescent systems in $0.4\leqslant z<0.6$ redshift range because of the magnitude limit. Thus we limit the analysis to the $0.1< z<0.4$ redshift range where the high density of hCOSMOS+zCOSMOS red galaxies provides a robust set of tracers for the quiescent galaxy stellar mass field.

\section{Smoothed Galaxy Stellar Mass Density Estimation}\label{estimation}

\subsection{The Method}\label{method}

\begin{figure}
\begin{centering}
\hspace*{-0.4in}
\includegraphics[scale=0.25]{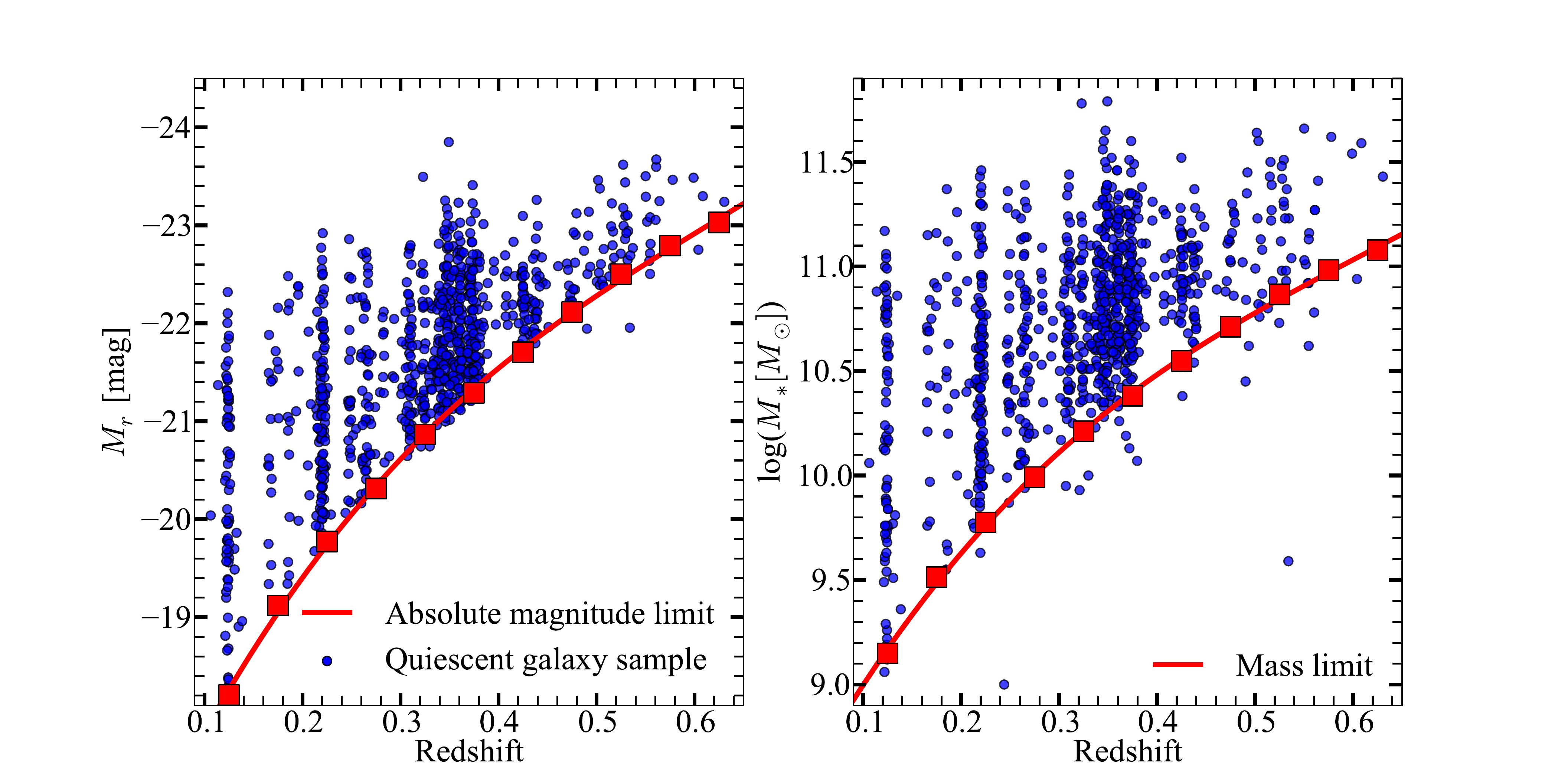}
\caption{Absolute $r-$band magnitude vs. redshift (left) and stellar mass vs. redshift (right). The blue points show parent spectroscopic sample of quiescent galaxies. Red curves and squares denote the redshift evolution of the absolute magnitude limit (left) and galaxy stellar mass limit (right) corresponding to the magnitude limit of the survey ($r_{lim}=20.8$). \label{f6}}
\end{centering}
\end{figure}

We follow a standard approach for evaluating the environments of compact galaxies \citep{Park1994, Grogin1998, Muldrew2012}. In particular we specialize to the procedures outlined by \citet{Tempel2012}.

To evaluate the galaxy stellar mass density field (GSMD) around quiescent galaxies we begin by constructing a continuous GSMD field. We identify 880 quiescent galaxies covering stellar mass range between $10^9\, M_\sun$ and $6.3\times10^{11}\, M_\sun$. We adopt a correction factor \citep[see e.g.,][]{Grogin1998, Tempel2012} to account for the unsampled end of the mass function. We transform the corrected point distribution of galaxy stellar masses in our magnitude-limited sample into a continuous  GSMD field.

In a magnitude limited sample, the fraction of galaxies more luminous than some fixed minimum luminosity naturally increases with redshift. Here we are interested in the GSMD rather than the luminosity density. Because we select quiescent galaxies, translating from one limit to the other is a reasonable approximation.

We assume that the galaxy stellar mass function is independent of compactness. We use the galaxy stellar mass function from \citet{Muzzin2013b}; this galaxy stellar mass function is derived for the redshift range $0.2 < z < 0.5$ of the COSMOS survey. Thus assuming this form is internally consistent.  

We use the galaxy stellar mass function to assign a redshift-dependent weight to each galaxy, $W_z$, that compensates for the incompleteness:

\begin{equation}
W_ z=\frac{\int_{M_l}^{M_u}M_\star \Phi\left(M_\star\right)\mathrm{d}M_\star}{\int_{M_{lim}(z)}^{M_u}M_\star \Phi\left(M_\star\right)\mathrm{d}M_\star} \label{eq:weight},
\end{equation}

\noindent where $\Phi\left(M_\star\right)$ is: 

\begin{equation}
\Phi(M) = \left(\ln 10\right)\Phi^{*}\left[10^{\left(M - M^{*}\right)\left(1+\alpha\right)}\right] \times \exp\left[-10^{\left(M - M^{*}\right)}\right]. \label{eq:mf}
\end{equation} 
       
\noindent Here $M=\log(M_\star/M_\sun)$ is galaxy stellar mass, $\alpha=0.92$ is the slope at low mass, $M^{*}=\log(M_\star^{*}/M_\sun)=11.25$ is the characteristic mass, and $\Phi^{*}$ is the normalization of the stellar mass function (irrelevant here). This galaxy stellar mass function is based on galaxy stellar masses from the COSMOS/UltraVISTA catalog \citep[][Section~\ref{data}]{Muzzin2013a}.

For the lower and upper stellar mass cuts we use the limiting galaxy masses we sample (see Section~\ref{mass}): $M_l=\log(M_\star^{l}/M_\sun)=9$ and $M_u=\log(M_\star^{u}/M_\sun)=11.8$. Because of the approximate constancy of the mass-to-light ratio $(M/L)_r$ with absolute magnitude for quiescent galaxies with D$_\mathrm{n}4000\gtrsim1.4$ \citep[e.g.,][]{Geller2014} the effective galaxy stellar mass limit at the galaxy redshift $M_{lim}(z)$ is directly related to the absolute magnitude limit $M_{r,lim}(z)$.  

We determine the absolute $r-$band magnitude at the observed redshift by synthesizing photometry in this band from the measured multi-band UV--IR spectral energy distribution (SED). We determine the $K-$corrected and reddening corrected magnitude by fitting stellar population synthesis models of \citet{Bruzual2003} to the observed SED using the LePHARE code \citep{Arnouts1999, Ilbert2006}. The absolute magnitude limit (red solid line in the left panel of Figure~\ref{f6}) is

\begin{equation}
M_{r,lim}(z)=m_{r,lim}-5\log\left(\frac{D_L(z)}{10\mathrm{pc}}\right)-\widetilde{K}\left(z\right), \label{eq:magl}
\end{equation}

\noindent where $m_{r,lim}=20.8$ is the apparent magnitude limit, $D_L(z)$ is the luminosity distance, and $\widetilde{K}(z)$ is the median $K-$correction.  

At each redshift we translate the magnitude limit $M_{r,lim}(z)$ into the galaxy stellar mass $M_{lim}(z)$ (red solid line in the right panel of Figure~\ref{f6}) using $(M/L)_r\sim1$, based on the approximately constant $(M/L)_r$ value for systems with D$_\mathrm{n}4000\gtrsim1.4$ in our hCOSMOS+zCOSMOS quiescent sample. Only 32 galaxies (3.6\% of the sample) have stellar masses below this galaxy stellar mass limit. 

Values of the redshift-dependent weighting factor $W_z$ are not large: they range between $1.00$ and $1.16$ for galaxies in the $0.1< z<0.4$ redshift interval.  The estimated effective total stellar mass per  galaxy from the magnitude-limited sample is then

\begin{equation}
M_\star^{tot}=M_\star\times W_z. \label{eq:mtot}
\end{equation}

\noindent The weighted stellar masses are the basis for the smoothed GSMD field.

We smooth the region around each galaxy in the sample with a $B_3$ spline kernel \citep{Tempel2012}:

\begin{equation}
B_3(x)=\frac{|x-2|^3-4|x-1|^3+6|x|^3-4|x+1|^3+|x+2|^3}{12}, \label{eq:b3}
\end{equation}

\noindent where $x=|\boldsymbol{r}-\boldsymbol{r}_{i}|/a$ is the distance between a given point $\boldsymbol{r}\equiv[\alpha, \delta, r(z)]$ and the $i-$th galaxy, normalized by the smoothing scale $a$. The GSMD field at each position is a sum over all galaxies in the sample:

\begin{equation}
m=\frac{1}{a^3}\sum_{i=0}^{N}B_3\left(\frac{|\boldsymbol{r}-\boldsymbol{r}_{i}|}{a}\right)M_\star^{tot},\label{eq:denfield}
\end{equation}

\noindent where $B_3(|\boldsymbol{r}-\boldsymbol{r}_{i}|/a)=0$ for $|\boldsymbol{r}-\boldsymbol{r}_{i}|\geqslant2a$. The number of positions where we calculate the GSMD field is equal to the number of galaxies in the sample. In order to avoid overweighting each galaxy with the spline, we sample the effective GSMD field at randomly chosen points around each object. We require that both the offsets from these random positions and the fiducial galaxy lie within the same cube of side length 1~Mpc. In effect, we calculate the GSMD field in non-empty cells of a cartesian grid. This analysis produces a continuous distribution of GSMD that we use to compare the GSMD field around massive compact systems with the GSMD field around the underlying quiescent population. 

\subsection{The Galaxy Stellar Mass Density Field Around Compact Quiescent Galaxies}\label{densityfield}

\begin{figure}
\begin{centering}
\hspace*{-0.3in}
\includegraphics[scale=0.4]{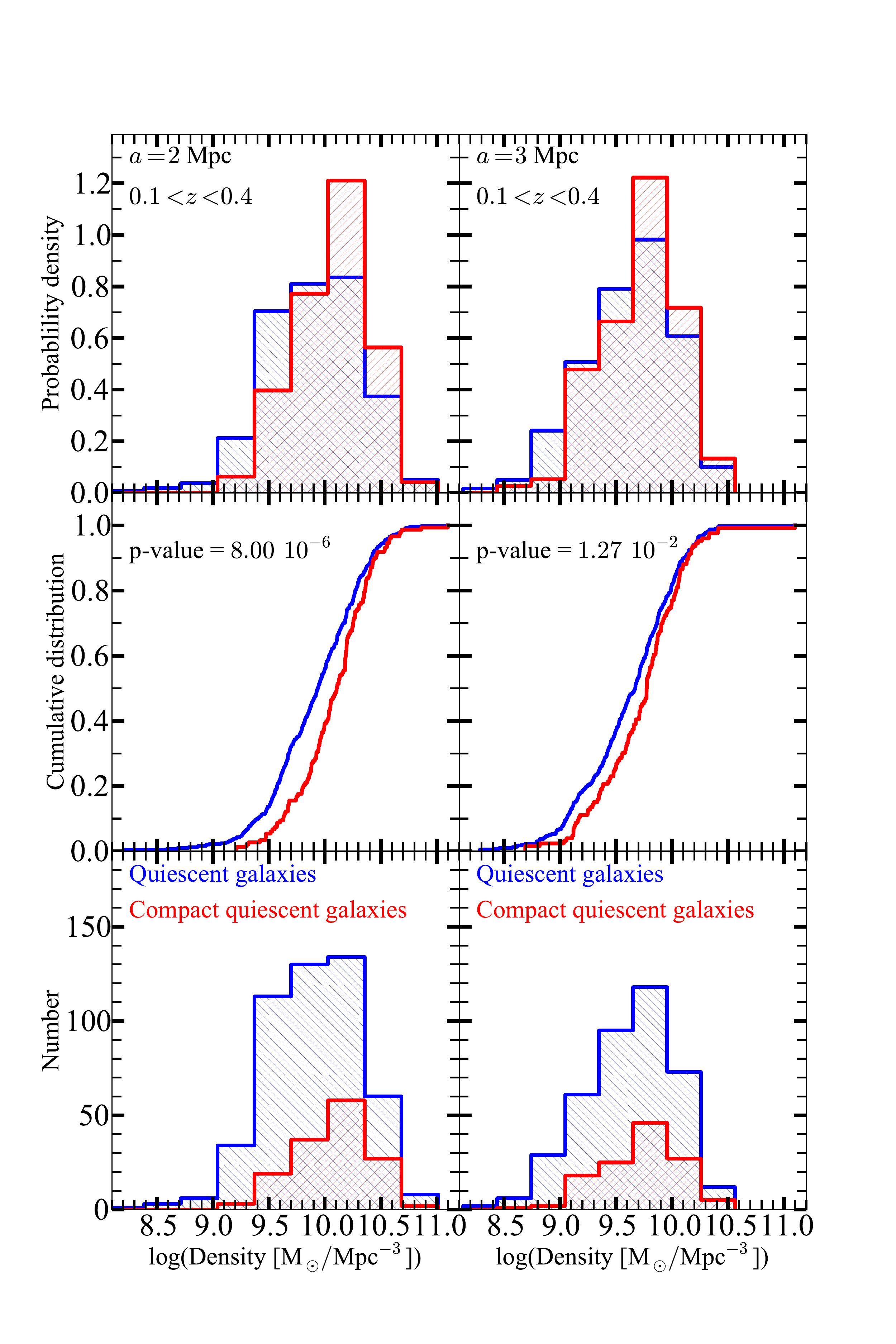}
\caption{Distribution of the GSMD around galaxies in the parent quiescent sample (blue) and for the quiescent compact subsample (red) sampled on a scale of 2 Mpc (left column) and 3 Mpc (right column). Top histograms show normalized probability densities. Central panels show cumulative distributions with the corresponding $p-$ values of the two-sample Anderson-Darling (A-D) test. The number distribution of galaxies in bins of top histograms are in the bottom panels. For both scales the two density distributions are inconsistent with being drawn from the same parent population, as indicated by the small $p-$values. \label{f7}}
\end{centering}
\end{figure}

\begin{figure*}
\begin{centering}
\hspace*{-0.5in}
\includegraphics[scale=0.35]{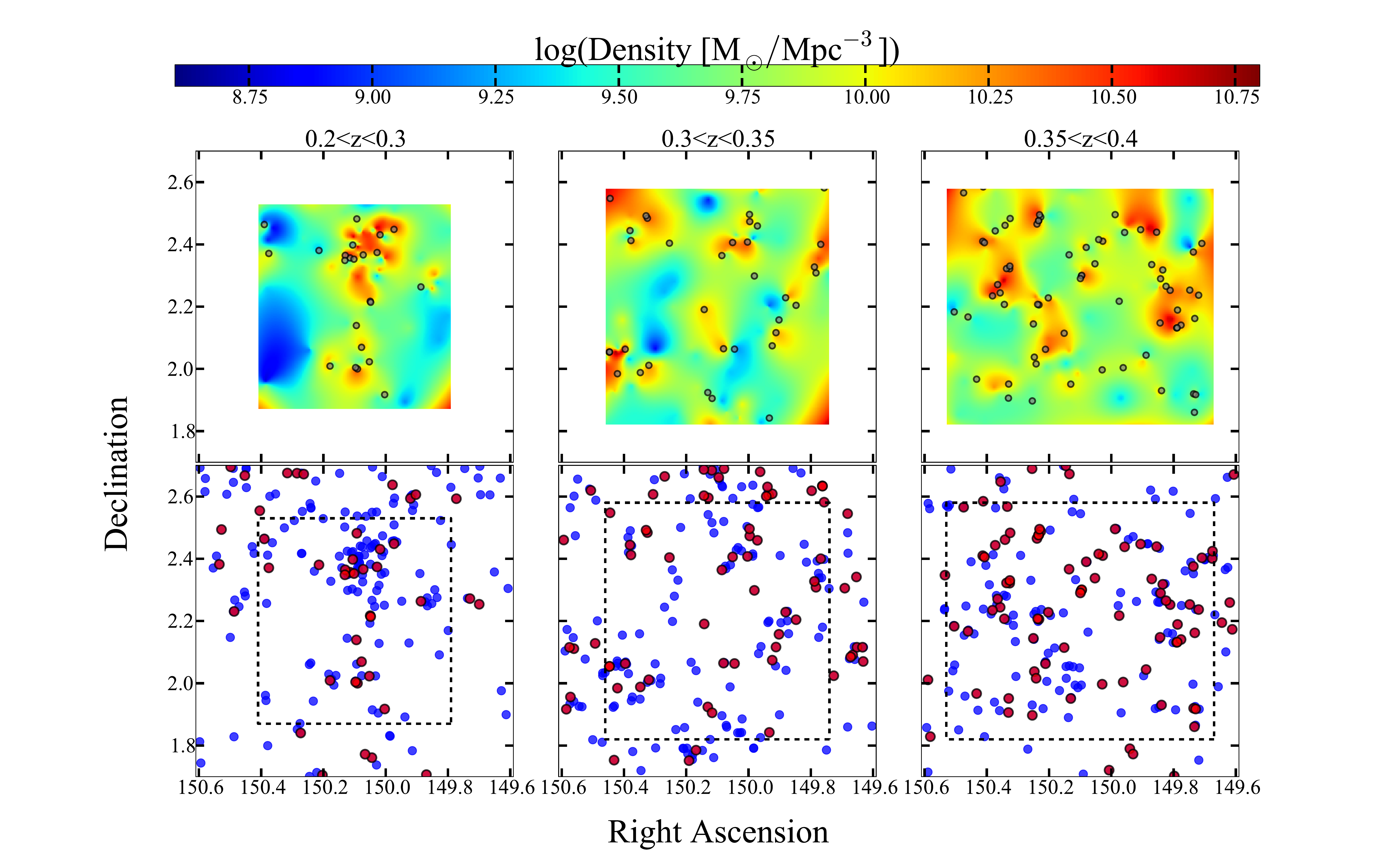}
\caption{Top panels: GSMD map for the smoothing length $a=2$~Mpc in three redshift slices encompassing similar volumes. Gray circles denote the positions of compact quiescent galaxies. Bottom panels: Spatial distributions on the sky of the parent quiescent sample (blue circles) and the massive compact subsample (red circles). It is visually evident that the compact galaxies are generally in denser regions.\label{f8}}
\end{centering}
\end{figure*}

We calculate the GSMD field for two smoothing lengths, 2~Mpc and 3~Mpc. The smoothing scales we explore are limited by: 1) the density of our targets in redshift space, and 2) the transverse dimension of the 1 sq. degree field at the limits of the redshift range we probe. 

As discussed in Section~\ref{comp}, the sampling of redshift space for $z \geq0.4$  is sparse. Environment determinations on the scales we can explore at lower redshift become dominated by shot noise because the mean galaxy separation,  $8-10$~Mpc at $z>0.4$, substantially exceeds even the $3$~Mpc smoothing length \citep{Grogin1998}. 

The mean galaxy separation at the peak of the redshift distribution  ($z\sim0.34$) is $\sim5$~Mpc. Simulations show that the reconstruction of over-dense regions fails for smoothing lengths much smaller than the mean galaxy separation in a densely populated redshift bin \citep{Cucciati2006}. Thus we select 2~Mpc as the smallest smoothing scale for  environment density estimation. At the scale of 2~Mpc $\sim75\%$~( 542/734) of the bins contain more than one galaxy.

The largest smoothing scale we probe is set by the transverse dimension of the field at the lower redshift limit of our sample. The central 1 sq. degree of the COSMOS field covers a scale $\gtrsim6$~Mpc for $z\sim0.1$. The scales covered by the survey must be substantially larger than the smoothing length to avoid edge effects. Thus  we select $3$~Mpc as the upper cut-off for the GSMD field smoothing scale; it is the largest scale where we can use the full extent of the redshift range. 

The GSMD field near the edge of the survey is underestimated if the smoothing length covers a region outside the survey area  \citep[e.g.,][]{Tempel2012}. To account for the edge effects we estimate the GSMD only in sections of the survey area that are farther than a smoothing length from survey edges.  For a given smoothing scale, the fraction of the survey volume excluded from the density estimation is a strong function of redshift. At $z\sim0.1$ and for $a=3$~Mpc we include only the central $\sim10\%$ of the survey. This fraction rises to $\sim55\%$ at $z\sim0.2$  for the same smoothing length. By considering only objects far enough from the survey boundaries we obtain an unbiased estimation of the GSMD field.

Figure~\ref{f7} shows the distribution of the GSMD field around quiescent COSMOS galaxies at $0.1< z<0.4$ (blue histograms) compared with the GSMD field surrounding massive compact systems (red histograms). The distribution of the GSMD field around massive compact galaxies shifts toward the high-density end of the distribution for the parent quiescent sample (top panels of Figure~\ref{f7}). This shift holds for both smoothing lengths. 

The two-sample Anderson-Darling (A-D) test \citep{Scholz1987}, based on the cumulative distributions in the two central panels of Figure~\ref{f7}, corroborates the visually apparent difference between the two sets of histograms. Difference between cumulative distributions is much larger for $a=2$~Mpc smoothing length. The probability that the two distributions are drawn from the same underlying distribution is $p=8.\times10^{-6}$ for the 2~Mpc scale. Thus for the smaller smoothing length the null hypothesis is rejected at a $4.3\, \sigma$ significance level. The probability is lower for 3~Mpc scale: $p=1.27\times10^{-2}$ gives a marginal $2.24\, \sigma$ rejection. This lower significance of the 3~Mpc result reflects the limitations of our 1 sq. degree magnitude-limited ($r<20.8$) survey. 

The number of quiescent and compact quiescent galaxies in the majority of the histogram bins in Figure~\ref{f7} exceeds $n=10$. Numbers are lower only for the two most extreme-density bins (bottom panels of Figure~\ref{f7}). At 2~Mpc the resulting contrast between the GSMD distributions for the compact and for the parent quiescent samples is thus not affected by small-number statistics.   

The top panels of Figure~\ref{f8} show maps of the GSMD field, constructed using a smoothing scale of 2~Mpc, in three redshift slices (of similar volumes) between $z=0.2$ and $z=0.4$.  We derive these density maps based on the spatial distribution of quiescent galaxies (blue circles in the bottom panels of Figure~\ref{f8}). To construct GSMD maps we use a linear radial basis function to interpolate between the positions where we evaluate the GSMD. We avoid the edges of the survey (areas outside dashed-line squares in Figure~\ref{f8}) because the GSMD is underestimated in regions where our sample is not complete. In each redshift slice, massive compact galaxies (gray circles in the top panels and red circles in the bottom panels of Figure~\ref{f8}) favor denser regions (see also Figure~\ref{f7}). 

We also examine the GSMD distribution for massive compact quiescent galaxy samples selected based on more conservative compactness thresholds. All compact samples exhibit the same preference for denser regions. However, a smaller number of systems in the samples with more extreme properties decreases the significance of the two-sample A-D test results. For the compact sample of 50 quiescent galaxies with $\Sigma_{1.5}>10.5$ (Eq.~\ref{eq:def}), the null hypothesis that the GSMD distributions for the compact and for the parent quiescent sample originate from the same underlying distribution is rejected at a marginal $2.3\, \sigma$ significance level ($p=1.07\times10^{-2}$). The sample of massive quiescent compact systems with $\Sigma_{1.5}>10.7$ contains only 10 objects; obviously inadequate for statistical analysis. A study of the environments of intermediate-redshift galaxy samples covering a broad range of compactness levels requires a survey covering a field larger than the COSMOS field.

\section{Interpretation: The Impact of Galaxy Stellar Mass}\label{mass}

\begin{figure}
\begin{centering}
%\hspace*{-0.325in}
\includegraphics[scale=0.5]{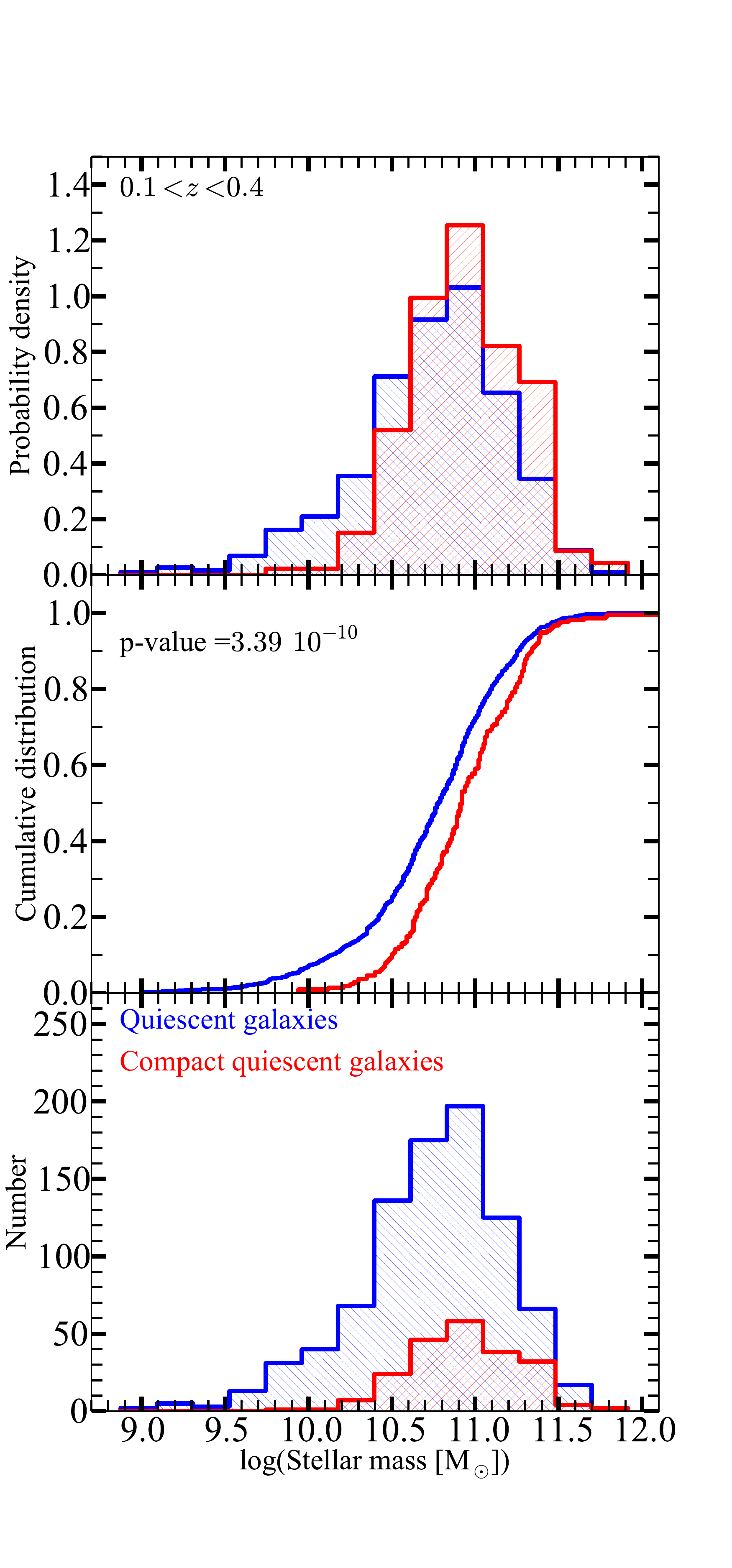}
\caption{Distributions of galaxy stellar masses for 1) the parent quiescent galaxy sample (blue) and 2) the compact subsample (red). The first panel show normalized probability densities. The second panel shows two cumulative distributions. The distributions of absolute galaxy numbers are in the third panel. The compact quiescent galaxies occupy the high-mass end of the parent stellar mass distribution.\label{f9}}
\end{centering}
\end{figure}

The denser environments of massive compact galaxies may be related to their stellar mass and/or to their dense internal structure (i.e, a combination of their mass and size). Here we explore the impact of the massive nature of compact galaxies. We use  magnitude-limited samples of non-compact quiescent galaxies and compact quiescent galaxies with the same stellar mass distribution to test whether the distribution of the GSMD around compact and non-compact objects differ. 

At $z<1$ massive ($M_\star>5\times10^{10}\, M_\star$) red galaxies, regardless of compactness,  reside preferentially in high-density regions (e.g., \citealt{Bolzonella2010, Darvish2015} (COSMOS), \citealt{Mortlock2014} (CANDELS - UDS and GOODS-S)). Thus the difference in the stellar mass distribution of our parent quiescent sample and the compact subsample may account for the observed difference in environments (Figure~\ref{f6}). 

Massive compact galaxies populate the higher-mass end of the distribution for the underlying quiescent population (Figure~\ref{f9}). The parent quiescent galaxy sample spans the galaxy stellar mass range from  $10^{9}\, M_\sun$ to $6.3\times10^{11}\, M_\sun$, with a median value of $\widetilde{M}_\star=6\times10^{10}\, M_\sun$. The minimum stellar mass for the compact subsample is $M_\star=10^{10}\, M_\sun$, and the median mass is $\widetilde{M}_\star=10^{11}\, M_\sun$. The two stellar mass distributions differ: a $p-$ value of $3.39\times10^{-6}$ corresponds to a $6.2\, \sigma$ rejection of the hypothesis that the two stellar mass distribution originate from the same parent sample (central panel of Figure~\ref{f9}).

To further investigate the relationship between the high stellar mass of compact galaxies and their surrounding GSMD, we use the compact stellar mass probability distribution (red hatched area in the top panel of Figure~\ref{f9}) to select a sample of non-compact ($\Sigma_{1.5}<10.3\, \log(M_\sun\, \mathrm{kpc}^{-1.5})$, see Eq.~\ref{eq:def}) quiescent objects with the same stellar mass distribution as the compact sample. We then evaluate the GSMD field around the most massive non-compact quiescent systems using the smoothing kernel density estimation technique described in Section~\ref{method}.

\begin{figure}
\begin{centering}
%\hspace*{-0.3in}
\includegraphics[scale=0.495]{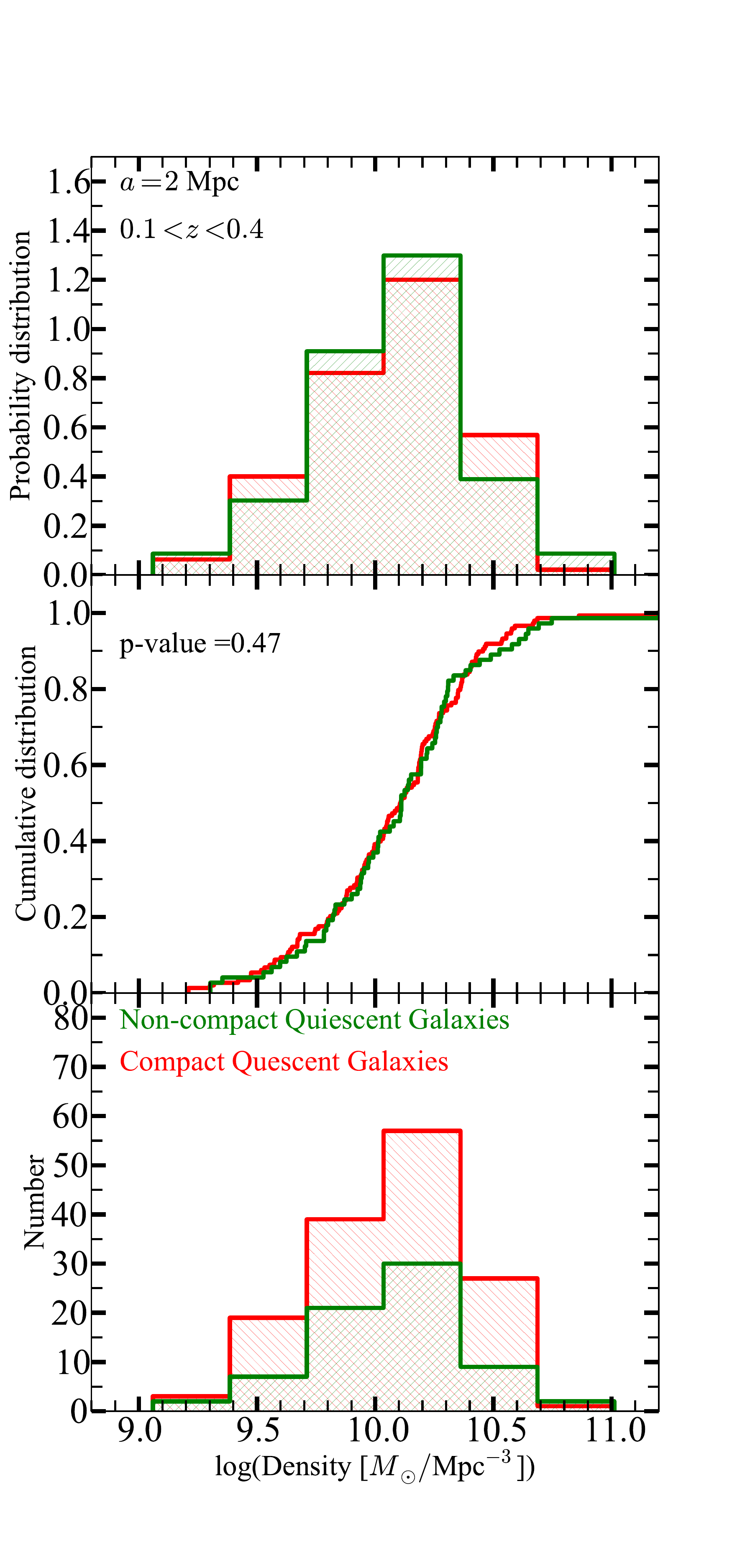}
\caption{Distributions of GSMD at a 2~Mpc smoothing scale for: 1) a randomly selected subsample of non-compact quiescent galaxies with the same stellar mass distribution as the compact galaxy subsample (green), and 2) the quiescent compact subsample (red, as in Figure~\ref{f6}). The top histograms show the normalized probability density; the corresponding numbers of galaxies are in the bottom histograms.  The central panel shows cumulative density distribution for the two subsamples. The GSMD distribution for the massive non-compact subsample is indistinguishable from the equivalent distribution for compact galaxies.
\label{f10}}
\end{centering}
\end{figure}

In Figure~\ref{f10} we compare the GSMD distribution around massive compact (red histogram) and non-compact objects (green histogram) for the 2~Mpc smoothing length (see Section~\ref{densityfield}). The distributions for the two types of objects are indistinguishable ($p\sim0.5$, central panel of Figure~\ref{f10}). This comparison suggests that the observed tendency of massive compact galaxies at $0.1<z<0.4$ in the COSMOS sample towards regions of greater surrounding GSMD is mainly driven by the greater stellar masses of compact systems.  

If the environment were strongly related to compactness rather than or in addition to the mass, we might expect a residual difference between the distributions presented in three panels of Figure~\ref{f10}. Much larger samples might enable detection of a difference between similarly constructed distributions. 

\section{Comparison With Previous Work}\label{comparison}

Several recent studies investigate the  relationship between massive ($M_\star>10^{10}\, M_\sun$) galaxy size and environment over cosmic time, with diverse results. Some of the results at high redshift ($z\gtrsim1$) suggest that quiescent color-selected galaxies in clusters on average have larger sizes than their massive analogs in less dense regions \citep{Cooper2012, Papovich2012, Delaye2014}. A similar difference apparently occurs between quiescent galaxy sizes in the highest and the lowest density regions \citep{Lani2013}. However, other high-redshift studies suggest the absence of any trend with environment \citep{vanderWel2008, Rettura2010, Newman2014}. Still others find that quiescent galaxies with early-type morphologies appear more compact in clusters than in lower density regions \citep{Strazzullo2010, Raichoor2012}.

Depending on the sample selection and definition of the environment, previous analyses of  intermediate-redshift quiescent galaxy samples paint an equally confusing picture.  Both galaxy counts in a fixed physical aperture and the distance to the nearest n-th neighbor suggest that at $0.5<z<1$ quiescent galaxies are larger in high local galaxy density environments. This positive trend in size with local galaxy density is less pronounced than at $z>1$ \citep{Lani2013}.  Group galaxies in the COSMOS field at $0.2<z<1$ selected either based on red rest-frame colors or on early-type morphology follow the same size-stellar mass relation as their counterparts in less dense regions \citep{Huertas-Company2013a}.  Sizes of massive early-type galaxies in the EDisCS survey at $0.4<z<0.8$ appear to be independent  of the environment \citep{Kelkar2015}. The authors find no trend in size with environment for a quiescent intermediate-redshift sample selected based on red rest-frame $B-V$ color. 

At $z\sim0$ the results are also inconclusive. Early-type galaxies appear to  follow the same size-mass relation in all environments \citep{Maltby2010, Huertas-Company2013b}. However, cluster galaxies in the PM2GC and WINGS survey display a different trend: quiescent dense galaxies are proportionally more numerous in clusters than in other types of environments \citep{Valentinuzzi2010, Poggianti2013b}. A large $z<0.12$ dataset drawn from SDSS DR7 shows that massive ($M_\star>4\times10^{10}\, M_\sun$) early-type galaxies are smaller in the over-dense than in under-dense regions \citep{Cebrian2014}, although the difference in size is not as pronounced as in PM2GC- and WINGS-based studies. 

To understand the range of results it is important to distinguish the impact of size from the impact of mass. In order to make this distinction, samples must be selected with identical mass distributions (Section~\ref{mass}). The well-known correlation between galaxy intrinsic luminosity and/or stellar mass and local density requires this approach \citep{Park2007, Scodeggio2009, Bolzonella2010}. It is unclear whether other studies take this known dependence into account.

Once the mass is taken into account,  the variety of results may persist because of  1) the use of photometric redshifts, 2) small, incomplete galaxy samples (especially at high redshift),  3) the range of environments probed, 4) and the techniques used for GSMD field estimation. Thus a direct quantitative comparison among these studies is not possible.

Other observational results show a trend toward smaller sizes of  $z<0.12$ massive early-type galaxies in overdense regions in SDSS \citep{Cebrian2014} and in nearby clusters \citep{Valentinuzzi2010, Poggianti2013b}. \citet{Wellons2015b} indicate that their theoretical results are also consistent with \citet{Valentinuzzi2010}. Results presented here may be consistent with the observed trends in low-redshift clusters \citep[and thus in agreement with some theoretical predictions of][]{Wellons2015b}, but our sample is not large enough to discriminate between the impact of mass and compactness.   

A concern in comparing the observations with theory is that the simulations \citep[e.g.,][]{Wellons2015a} predict a steep decline in the number density of massive compact quiescent systems with redshift, discrepant with the most recent observational evidence \citep[e.g.,][]{Poggianti2013a, Carollo2013, Damjanov2014, Damjanov2015, Saulder2015, Tortora2015}. It is noteworthy that the observations indicating a much milder decline in the abundance of massive compact quiescent systems cover large volumes \citep[e.g.,][]{Damjanov2014,Saulder2015,Tortora2015}, much larger than the simulations \citep{Wellons2015a}. Resolution of these issues and a more extensive consideration of the relations among galaxy compactness, mass, and environment are an impetus for surveys covering larger volumes than offered by COSMOS. 

%These observational and theoretical results may be consistent with our results,} but our sample is not large enough to discriminate between the impact of mass and compactness. In our sample the stellar mass distribution of the compact sample alone accounts for the shift toward denser environments.

\section{Conclusions}\label{sum}

The abundance of massive compact galaxies at intermediate redshift enables investigation of the relation between the internal properties of these systems and their environments based on a dense spectroscopic survey. We conduct a spectroscopic survey of the central 1 sq. degree region of the COSMOS field (hCOSMOS, Damjanov et al. 2015 in prep.) to construct 90\% complete sample to the limiting magnitude $r_{lim}=20.8$. 

We use the redshift survey to select a complete magnitude limited sample of quiescent galaxies based on the combination of $U-V$ and $V-J$ rest-frame colors.  The D$_\mathrm{n}4000>1.44$ spectroscopic indicator of galaxy quiescence agrees well with the rest-frame $UVJ$ selection.  

We estimate the GSMD for two smoothing lengths (2 and 3~Mpc) around quiescent systems at $0.1<z<0.4$. Using the compactness definition from \citet{Barro2013}, we select a sample of massive compact quiescent systems and compare their environments with the GSMD field of the parent quiescent population.  

Massive quiescent compact systems prefer denser regions (Figure~\ref{f8}).  At a 2~Mpc smoothing scale the GSMD for massive compacts is significantly offset toward higher-density relative to the one for the complete quiescent sample. The difference in the stellar mass distributions between compact and non-compact quiescent galaxies completely accounts for this shift. 

The massive nature of quiescent compact galaxies appears to be fundamental to their formation and evolution. There may be a further dependence of the size of the objects on environment, but there is no detectable signal of this additional effect in our sample. Much larger samples might enable exploration of the possibly subtle impact of the size. A survey over a larger area would enable exploration of a broader range of smoothing scales.

\acknowledgments
We thank the referee for providing comments that helped us clarify points in the paper. We thank  Warren Brown, Scott Kenyon, and Kate Rubin for reading the manuscript and making insightful suggestions. ID is supported by the Harvard College Observatory Menzel Fellowship. The Smithsonian Institution supports the research of MJG. HJZ gratefully acknowledges the generous support of the Clay Fellowship.
 
%\bibliography{References}

\end{document}